***Peer Review at Capacity: An editor's view***

Yamir Moreno[1,2,*],
[*]Editor-in-Chief, Journal of Complex Networks
[1]Institute for Biocomputation and Physics of Complex Systems, University of Zaragoza, Zaragoza 50018, Spain.
[2]Department of Theoretical Physics, Faculty of Sciences, University of Zaragoza, Zaragoza 50009, Spain.

**Abstract**: From the editor's desk, the pressure on peer review is no longer an abstract concern. It appears in the growing number of invitations needed to secure an appropriate report, the repeated reliance on a small group of dependable colleagues, and the duplication of work when manuscripts move between journals. The problem is not simply that science has too few potential reviewers. It is that reviewing labor is unevenly distributed, weakly recognized, and organized through a publication architecture built for a smaller system.

**A view from the editor's desk**

Scientific publishing has changed scale. The number of articles indexed in major databases was approximately 47% higher in 2022 than in 2016 **[1]**, while the longer-term literature has grown at an average rate of about 4% per year **[2]**. This is not in itself a failure. It reflects the globalization of research and lower barriers to communicating results. Yet the machinery used to evaluate those results has not adapted at the same pace.

As Editor-in-Chief of the Journal of Complex Networks, I encounter this mismatch in a practical form. Finding reviewers has always required judgment and persistence. What feels increasingly fragile is the distance between the number of people who could, in principle, review a paper and the much smaller number who possess the right combination of expertise, independence, availability, and willingness to provide a careful report within a reasonable time.

The difficulty is especially visible in an interdisciplinary journal. A paper may combine network theory, statistical physics, applied mathematics, computer science, and a substantive empirical domain. No single reviewer can assess every component with equal authority. Editors must assemble complementary expertise across communities that may use different standards of evidence. A manuscript that is easy to classify can be difficult to evaluate well.

These reflections are personal, but the underlying burden is measurable. One estimate concluded that researchers devoted more than 100 million hours to peer review in 2020 **[3]**. The figure is approximate, but its scale is revealing: science already invests enormous expert labor in evaluation. The question is whether that labor is being allocated intelligently.

Readers of this journal will recognize the structure of the problem. Peer review is itself a network linking authors, editors, reviewers, journals, institutions, and fields. Its degree and load distributions are highly heterogeneous. A small group of visible and reliable researchers becomes a set of hubs to which invitations repeatedly flow. As their load increases, response rates decline, editorial searches lengthen, and local overload propagates through the system.

This is more than a metaphor. A study of the biomedical literature found that aggregate reviewing capacity could exceed demand under several assumptions, but that the work was strongly concentrated: 20% of researchers performed between 69% and 94% of reviews **[4]**. Evidence from one domain should not be generalized uncritically to all of science, but the distinction is important. The problem is not necessarily an absolute lack of qualified people. It is a failure of distribution, incentives, matching, and coordination. The incentive structure makes the imbalance predictable. Publications influence appointments, promotion, and funding; peer review is still treated largely as invisible service. Researchers are rewarded for generating another manuscript, but rarely for spending several hours improving someone else's.

When a paper is rejected and submitted elsewhere, its previous reports often disappear from the process. New reviewers repeat parts of an assessment that has already been performed, while the original reviewers receive little recognition and the authors lose time. Peer review has never guaranteed that a result is correct. Its indispensable contribution is more modest and more realistic: it exposes claims, methods, and interpretations to informed criticism before they become part of the certified literature. Under sustained pressure, however, this evaluation can become shallow. Editors may settle for available rather than ideal referees. Reports may focus on presentation while missing assumptions, methodological weaknesses, or the limits of an inference. Delay is visible; loss of depth is harder to measure, and perhaps more damaging. The costs are unevenly distributed. Early-career researchers may depend on publication timelines for jobs, fellowships, or grants. Scholars with heavy teaching loads have less time to review, even when they have relevant expertise. Junior researchers frequently contribute without being named or credited. Largely uncompensated and weakly documented labor can reproduce inequalities while appearing formally open.

Editors therefore have a responsibility not only to obtain reviews, but to use reviewer attention carefully. This includes rigorous and timely editorial triage. For a broad journal such as ours, technical correctness is necessary but not always sufficient to justify external review. A submission should also make a clear advance and offer mathematical, physical, methodological, or phenomenological insight of interest to the complex-networks community. A prompt editorial decision, with a concise and honest explanation, is often more respectful than sending a poorly matched paper into an already congested process.

Some improvements do not require a complete redesign of scholarly communication. Journals can broaden and diversify their reviewer pools, including greater participation by early-career researchers. This must be done transparently. Co-reviewing should be declared, the junior contributor should receive credit, and mentoring should not become a way of transferring hidden labor. Editors can also search beyond the most visible names, rotate invitations more deliberately, and keep better records of workload and expertise.

Reviewing should be recognized as an intellectual contribution rather than a simple number. A crude quota would invite superficial reports; what matters is the quality, timeliness, and usefulness of the assessment. Journals can provide verified records and feedback, while universities and funders can recognize high-quality reviewing and editorial work in narrative curricula vitae and promotion dossiers.

We should also make reviews more reusable. With author and reviewer consent, a report could follow a manuscript across versions or journals. A new editor would remain free to seek additional expertise or disregard an inadequate assessment, but should not be forced to behave as if earlier evaluation had never occurred. Portable review would reduce duplication and give careful reports a life beyond a single editorial decision. Finally, editorial procedures should help reviewers concentrate on scientific judgment. Routine checks for completeness, ethics statements, data availability, reference integrity, or reporting conventions should occur before a manuscript reaches a referee. Reviewers should spend their time where expertise matters most: whether the problem is important, the method is appropriate, the evidence supports the conclusion, and the contribution changes our understanding.

The larger reform is to separate functions that journals have traditionally bundled together: dissemination, evaluation, curation, certification, and archiving. Preprints already show that dissemination need not wait for journal acceptance. Their public status must be communicated clearly, but making a manuscript available and certifying its quality are different acts.

Evaluation could likewise be organized around the manuscript rather than the venue to which it happens to be submitted. The publish-review-curate model proposed by Stern and O'Shea **[5]**, together with initiatives such as Peer Community In **[6]**, provides working elements of this alternative. Journals would not disappear. Their role could become more intellectually distinctive: selecting, contextualizing, and certifying valuable work from a wider scholarly record instead of controlling the first moment at which the work can be seen.

Such a transition also depends on research assessment. DORA **[7]** and CoARA **[8]** have argued against using journal-level indicators as proxies for the quality of individual work. Unless hiring, promotion, and funding practices change accordingly, researchers will continue to face strong incentives to maximize publication counts and pursue journal brands, thereby feeding the submission pressure that journals are asked to absorb.

Artificial intelligence adds urgency and opportunity. Generative tools reduce the cost of producing fluent text and may increase submissions without increasing expert attention. The same technologies can assist with reference checking, reporting completeness, reviewer matching, and detecting internal inconsistencies **[9]**. Used carefully, they can remove routine work from editors and reviewers. They cannot, however, assume responsibility for scientific judgment. Models may reproduce historical biases, miss unconventional contributions, fabricate criticisms, or mishandle confidential material. Any use of AI in editorial or reviewing work should therefore be bounded, secure, and disclosed where appropriate. A human editor or reviewer must remain accountable for every substantive assessment. The purpose of automation should be to protect expert attention, not to create the appearance that expertise has been supplied.

Concluding, I do not believe peer review is obsolete, nor that it has already collapsed. I do believe that parts of the system are operating close to their practical limits. Its resilience still depends too heavily on the goodwill of a minority of researchers and on editors repeatedly solving the same allocation problem manuscript by manuscript. No journal can repair this architecture alone. Nevertheless, journals can act: triage responsibly, broaden participation, recognize reviewing, reuse existing evaluations, and reserve expert time for questions that require expert judgment. Publishers, repositories, funders, and institutions must then provide the interoperable infrastructure and incentives that allow these practices to scale. The peer-review system is a common scientific resource. Those of us who submit work also share a responsibility to evaluate the work of others, although reciprocity should be understood as a professional norm rather than a mechanical transaction. Network science teaches us that resilient systems distribute load, reduce avoidable bottlenecks, and preserve function under growth. Scientific publishing now needs to apply that lesson to itself.